\documentclass{jpp}
\usepackage{graphicx}
\usepackage{natbib}
\usepackage[T1]{fontenc}
\usepackage{bm}
\usepackage{color}
\usepackage{amsmath}
\usepackage{amssymb}
\usepackage{amsfonts}
\usepackage{courier}
\usepackage{txfonts}

\newcommand{\appref}[1]{\hyperref[#1]{Appendix~\ref{#1}}}

\usepackage{ae}
\usepackage{units}
\usepackage[disable]{todonotes}

\newcommand{\be}{\begin{displaymath}}
\newcommand{\ee}{\end{displaymath}}
\newcommand{\bn}{\begin{equation}}
\newcommand{\en}{\end{equation}}

\usepackage{enumerate}

\usepackage[T1]{fontenc}
\usepackage{bm}
\usepackage{color}
\usepackage{graphicx}
\usepackage[breaklinks=true]{hyperref}
\hypersetup{
  unicode=false,          % non-Latin characters in Acrobat’s bookmarks
  pdftoolbar=true,        % show Acrobat’s toolbar?
  pdfmenubar=true,        % show Acrobat’s menu?
  pdffitwindow=false,     % window fit to page when opened
  pdfstartview={FitH},    % fits the width of the page to the window
  pdfsubject={Subject},   % subject of the document
  pdfcreator={Creator},   % creator of the document
  pdfproducer={Producer}, % producer of the document
  pdfkeywords={keyword1} {key2} {key3}, % list of keywords
  pdfnewwindow=true,      % links in new PDF window
  colorlinks=true,        % false: boxed links; true: colored links
  linkcolor=blue,         % color of internal links (change box color with linkbordercolor)
  citecolor=blue,         % color of links to bibliography
  filecolor=blue,         % color of file links
  urlcolor=blue           % color of external links
}

\usepackage{amsmath}
 \usepackage{gensymb}

\title{Proton Acceleration in a Laser-induced Relativistic Electron Vortex}

\author{L.~Q.~Yi\aff{1}\corresp{\email{longqing@chalmers.se}}, I.~Pusztai\aff{1}, A.~Pukhov\aff{2}, B.~F.~Shen\aff{3}, \and T.~F\"ul\"op\aff{1}} 

\affiliation{\aff{1}Department of Physics, Chalmers University of Technology,
 SE-41296 G\"{o}teborg, Sweden
\aff{2}Institut f\"ur Theoretische Physik I, Heinrich-Heine-Universit\"at D\"usseldorf, D\"usseldorf, 40225 Germany  
\aff{3}Department of Physics, Shanghai Normal University, Shanghai, 200234, China}

\begin{document}

\maketitle

\begin{abstract}
 We show that when a solid plasma foil with a density gradient on the front surface is irradiated by an intense laser pulse at a grazing angle, $\sim 80^{\circ}$, a relativistic electron vortex is excited in the near-critical-density layer after the laser pulse depletion. The vortex structure and dynamics are studied using  particle-in-cell simulations. Due to the asymmetry introduced by nonuniform background density, the vortex drifts at a constant velocity, typically $0.2-0.3$ times the speed of light. The strong magnetic field inside the vortex leads to significant charge separation; in the corresponding electric field initially stationary protons can be captured and accelerated to twice the velocity of the vortex (100--200 MeV). A representative scenario---with laser intensity of 10$^{21}$ W/cm$^2$---is discussed: two dimensional simulations suggest that a quasi-monoenergetic proton beam can be obtained with a mean energy 140 MeV and an energy spread of $\sim 10\%$. We derive an analytical estimate for the vortex velocity in terms of laser and plasma parameters, demonstrating that the maximum proton energy can be controlled by the incidence angle of the laser and the plasma density gradient.
\end{abstract}

\section{Introduction}
\indent

Acceleration of protons by intense lasers has attracted extensive interest due to the potential significance in several branches of science, technology and medicine \citep{Daido2012}. The most experimentally robust mechanism for laser-driven proton acceleration is target normal sheath acceleration (TNSA) \citep{Wilks2001, Pukhov2001, Roth2002, Mora2003, Fuchs2006}, where protons are accelerated by the electrostatic sheath field arising due to the expansion of laser-heated electrons. It usually leads to a broad proton energy spectrum due to lack of control of the heating process \citep{Daido2012}, and the maximum achieved proton energy of 94 MeV \citep{Higginson2018}, is still not sufficiently high for many of the foreseen applications; e.g.~hadron therapy requires 100--200 MeV mono-energetic proton beams \citep{Bulanov2008}. In other proposed schemes, 
%such as collisionless shock acceleration \citep{Silva2004, Haberberger2012}, and radiation pressure acceleration [including hore-boring \citep{Badziak2004,Akli2008,Palmer2011,Schlegel2009} and light sail \citep{Esirkepov2004, Yan2008, Qiao2009, Henig2009, Kar2012, Kim2016}],
the acceleration usually involves a slow-moving acceleration structure (i.e.~such that protons can keep up with it) that can be driven by the piston action of the laser, radiation pressure, or a shock wave that is detached from the laser \citep{Macchi2013,Borghesi2016,Schreiber2016,Macchi2017}. The protons then stay in phase with this structure and get energized over a longer period of time, which typically leads to higher cut-off energies. However, in most cases the acceleration is strongly limited by the instabilities on the interface where the light pressure is balanced by the plasma fields \citep{Pegoraro2007, Palmer2012, Khudik2014}.

In this paper, we describe a new proton acceleration mechanism that relies on a laser-induced relativistic electron vortex (EV) in a near-critical density (NCD) plasma with a density gradient normal to the target surface [see Fig.~1(a)]. The EVs are not susceptible to the instabilities arising at the laser-plasma interface, as they form after the laser depletion.
Here the EVs are micro-scale, long lived, coherent magnetic flux ropes with an unipolar magnetic structure in 2D, and share similarities with a particular class of stable solutions to the electron magnetohydrodynamic equations \citep{Nycander1991, Gordeev1998, Yadav2008, Richardson2013, Angus2014}. Unlike the classic TNSA scheme, the acceleration mechanism described here is able to produce a quasi-monoenergetic proton beam, and the drift velocity of the EV---and thus the beam energy---can be tuned more easily compared to the collisionless shock acceleration \citep{Silva2004, Haberberger2012, Fiuza2012,Pak2018}. In addition to the application for proton acceleration, which is our main focus here, EVs have numerous potential applications. The strong coherent magnetic fields can be utilized in studying reconnection \citep{Yamada2010, Priest2003, Yi2018}, or the evolution of magnetic helicity in a kinetic system \citep{Wiegelmann2001, Rust1996}], thereby representing a fundamentally interesting setting to the wider plasma physics community.

Previous particle-in-cell (PIC) simulations have shown vortices in connection with a high intensity laser propagating through an underdense plasma \citep{Bulanov1996, Naumova2001}, but their potential for proton acceleration was not considered. In this work, we show that with a high-intensity laser at grazing incidence on a solid plasma foil, a strong relativistic EV can be excited in a NCD plasma layer after the laser depletion. Although the physics of laser pulse propagation in NCD plasmas has been widely studied \citep{Pukhov1996, Nakamura2008, Bulanov2010, Liu2013, Bin2015, Liu2016, Hilz2018, Ma2019}, the current study focuses a new regime, where the plasma density inhomogeneity length-scale is comparable to the vortex size in the direction normal to the surface of the target ($y$), while in the tangential directions, the density length-scale is much larger. Such density profile can be created with the pre-pulse of the laser evaporating the front surface of a solid plasma foil. Alternatively, when a sufficiently high-contrast is achieved \citep{Thaury2007}, the pre-plasma can be created by another laser with very large spot-size irradiating the target prior to the main pulse \citep{Gauthier2014}.
The acceleration mechanism considered here is different from the previously studied magnetic vortex acceleration (MVA) \citep{Bulanov2007, Nakamura2010, Bulanov2010, Lemos2012}, where bipolar magnetic vortex is generated in a uniform NCD target \citep{Nakamura2008} and acceleration only happens when it reaches the rear side of the target. In our work, the EV, or unipolar magnetic vortex, serves as a stable slow-moving structure in which protons can be captured, and gain considerable kinetic energy from the charge separation field associated with the EV. Similarly to collisionless shock acceleration, protons initially at rest can be reflected to twice the EV drift velocity \citep{Macchi2013}, and a narrow energy spectrum can be obtained in the direction of the vortex propagation.

\section{EV formation and the proton acceleration.}

\begin{figure}
\centering
\includegraphics[width=1\textwidth]{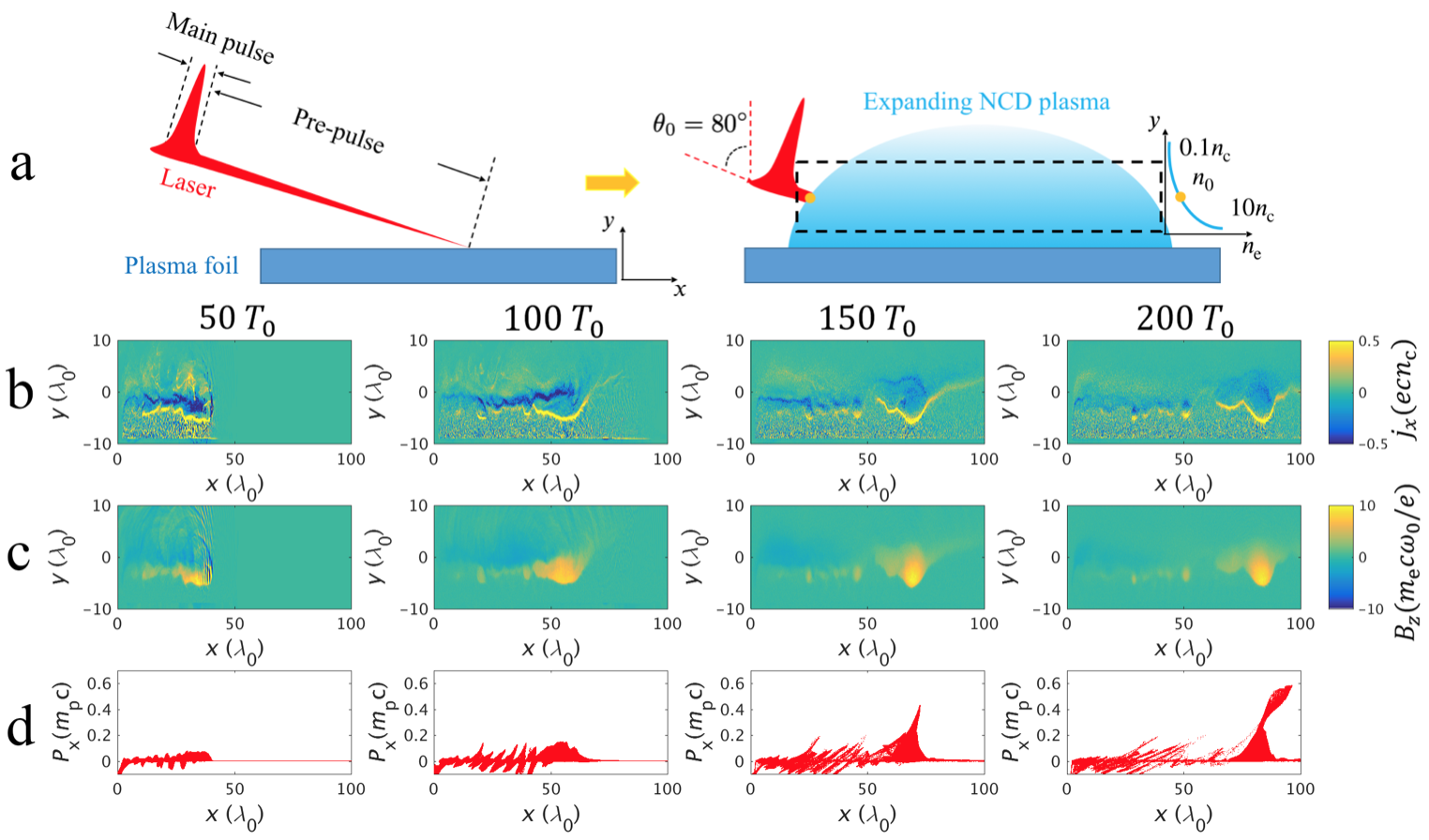}
\caption{Schematic of the proposed proton-acceleration setup. (a) A laser pulse grazing incident on a plasma foil (left). After heating by the laser pre-pulse, an expanding plasma region is formed on the surface (cyan shaded region), in which the main pulse interacts with the NCD layer (right). The black dashed box represents the simulation area, the orange dot marks the main pulse incident point, where the density is $n_0 \approx 0.4n_{\rm{c}}$. Snapshots of (b) the current density parallel to the target surface $j_x$, (c) the out-of-plane magnetic field $B_z$, and (d) the proton phase space for the $x$-direction, are shown at $t = 50T_{0}, 100T_{0}, 150T_{0}$ and $200T_{0}$, respectively.}
\label{fig.1}
\end{figure}

Figure~1 shows the simulation setup.  A linearly (p-)polarized laser beam with normalised amplitude $a_0 = 30$ (intensity $I = 1.2\times10^{21}$ W/cm$^{2}$), is incident at a grazing angle of 80$^{\circ}$ on a plasma foil. Here $a_0 = eE_0/m_{\rm{e}}c\omega_0$, $E_0$, $\omega_0$, and $\lambda_0  = 1\mu\rm m$ are the peak electric field, frequency, and wavelength of the laser pulse, while $c$, $e$ and $m_{\rm{e}}$ denote the speed of light in vacuum, the elementary charge and electron mass, respectively. The focal spot and the FWHM duration of the main laser pulse are $w_0=3\lambda_0$ and $5.4 T_0 = 18$ fs, respectively, where $T_0$ is the laser period. 
The main laser pulse propagates to the right with a grazing angle with respect to the target surface. It first penetrates the underdense plasma region until it reaches the simulation box (marked by the black-dashed line in Fig.~1, there is a 2-$\mu$m vacuum gap between the plasma and the left boundary where laser is injected), the density at the incident point is $n_0 \approx 0.4 n_c$ (in reality this is the position where the density becomes strongly inhomogeneous).
The density profile within the simulation box is initialised with a density decreasing exponentially in the $y$ direction as $n_{\rm{e}}(y) = 100n_{\rm{c}}\exp[-(y-y_{\rm{t}})/\sigma]$, where $n_{\rm{c}}=m_{\rm{e}}\omega_{0}^{2}/4\pi e^{2}$ is the critical density, $y_{\rm{t}} = -16.9 \lambda_0$ is the target position, and $\sigma = 3\lambda_0$ is the scale length. The density profiles in the tangential directions are assumed to be uniform since the scale-lengths in those directions are much larger than $\sigma$. Note that it is possible to achieve such density profile experimentally, as reported by \citet{Gauthier2014}. The proton density is $n_{\rm{p}}(y) = n_{\rm{e}}(y)/55$. 
The proton species is assumed to consist only a small fraction of the total positive charge, the majority of ions are considered to be immobile in this simulation. The effect of heavy ion motion will be addressed in the Discussion section. The PIC simulations are performed with EPOCH \citep{Arber2015}. The simulation window $x\times y= 200\lambda_{0} \times 20\lambda_{0}$ is sampled by $4000 \times 400$ cells with 10 macro electrons and 20 macro protons in each cell.

Figure~1(b-c) shows the evolution of the current density ($j_x$) and the out-of-plane magnetic field ($B_z$) during the interaction. The first column ($t = 50T_0$) shows the laser channeling in the NCD plasma, and the associated strong currents inside the channel due to the laser-plasma interaction. 
In the NCD plasma, the laser effectively transfers all its energy to the electrons within a distance of a few tens of microns \citep{Sylla2013}. In the second column ($t = 100T_0$), the laser energy is completely depleted and the electron dynamics is dominated by self-generated fields. The electrons gyrate in the self-generated magnetic fields, forming a stable vortex structure. The two columns on the right ($t = 150T_0$ and $200T_0$) show that the EV continues to propagate along $x$ with a drift velocity $\sim0.28 c$. In the meantime, the background protons are accelerated in the $x$-direction by the EV as shown in Fig.~1(d).

\begin{figure}
\centering
\includegraphics[width=1\textwidth]{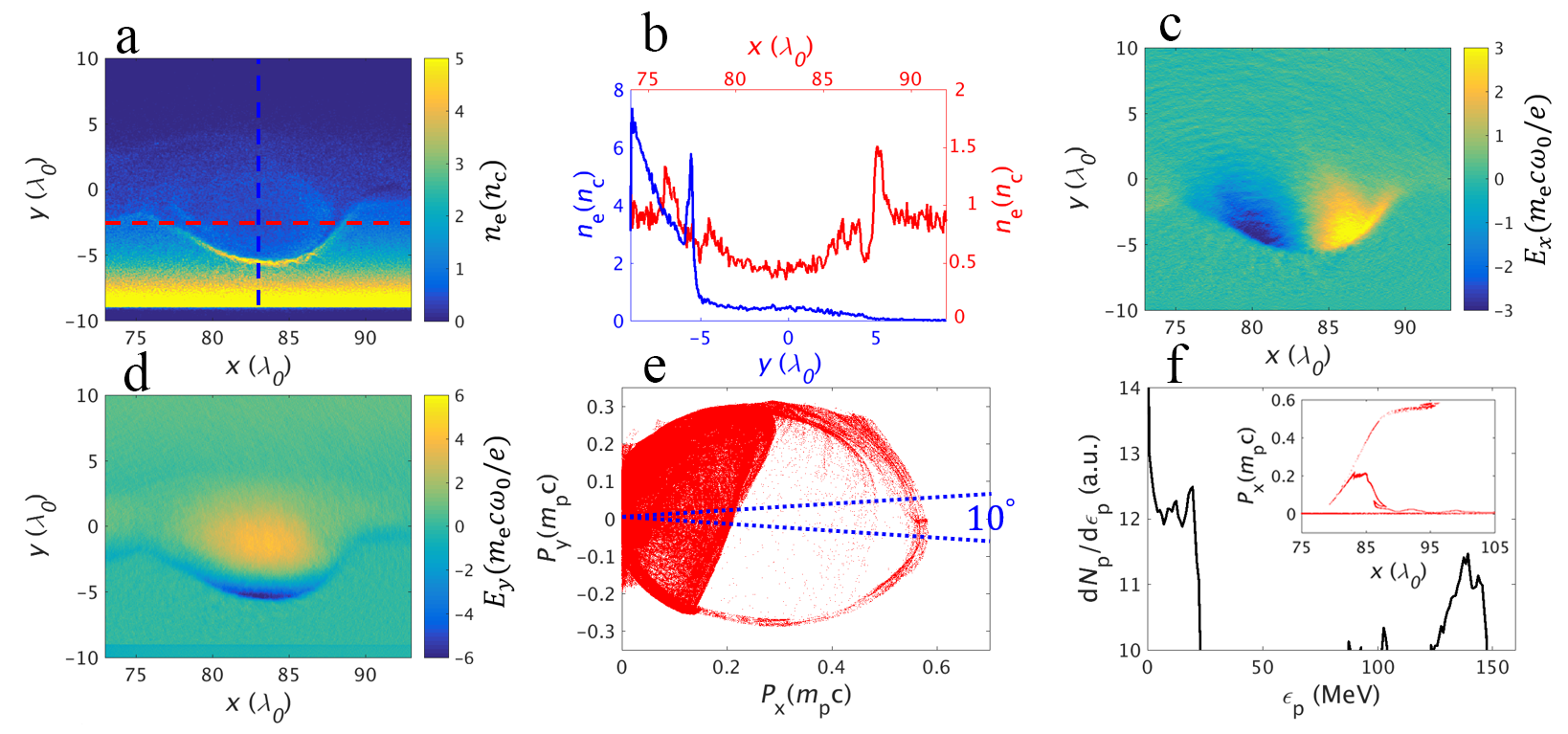}
\caption{Features of the relativistic EV. (a) The electron density distribution, where (b) presents the 1D cut of electron density along the horizontal (red) and vertical (blue) axes [marked as dashed in (a)]. Electric field components in the directions (c) parallel and (d) normal to the target surface. (e) Proton population in momentum-space $P_x$--$P_y$, and (f) energy spectrum of protons within opening angle 10$^{\circ}$ [between blue dashed lines in (e)]. The inset in (f) shows the phase space map $x$--$P_x$ of these protons. All quantities are shown at simulation time $t = 200T_0$.}
\label{fig.2}
\end{figure}

The acceleration of protons is due to the charge separation caused by the magnetic pressure in the EV. As shown by Fig.~2(a-b), the electron density inside the EV is lower than the ambient value. A high-density return current layer forms on the lower boundary of the EV. Components of the electrostatic field induced by the charge separation are presented in Fig.~2(c-d). 

As the EV drifts, the protons are accelerated in the leading half of the structure. Those protons that gain enough energy can keep up with the EV (they are \emph{captured}), otherwise they lose energy due to the decelerating field in the second half of the EV. The captured protons can be continuously accelerated until they overrun the vortex with approximately twice its drift velocity.

The electric field in the direction normal to the target surface ($E_y$) is defocusing, therefore only a fraction of the captured protons can achieve the maximum energy gain, others are scattered out of the acceleration phase before they can do so. This makes the energy distribution depend strongly on the pitch-angle as indicated by Fig.~2(e), where the slope of the dots shows the pitch-angle $\theta_{\rm{p}} = P_y/P_x$ of the corresponding protons. The energy spread of the accelerated protons can then be adjusted by selecting different pitch-angles (e.g.~with a collimator): at large $\theta_{\rm p}$ a wide spectrum is obtained with a sudden cut-off, while protons at low $\theta_{\rm p}$ exhibit a quasi-monoenergetic feature. External focusing with various well-established approaches \citep{Toncian2006, Albertazzi2015, Zhai2019} can be reliably employed in order to obtain a well-collimated beam for applications such as cancer therapy.

Figure~2(f) presents the energy spectrum of the protons within a 10$^{\circ}$ opening angle in the forward direction. The proton beam has a mean energy $\sim$140 MeV and an energy spread around 10$\%$. In this quasi-monoenergetic beam the proton number per 1$\%$ energy range and solid angle increment $\rm{\Delta}\Omega$ is estimated to be $\sim10^6$/msr$/1\%\epsilon_{\rm{p}}$  (assuming the proton beam to be axisymmetric with respect to the $x$ axis), which is comparable to contemporary laser-driven ion acceleration schemes at their cut-off energy \citep{Schreiber2016}, and this can be further improved by increasing the proton-to-heavy-ion ratio in the plasma.\\

In order to obtain a deeper insight into the electron dynamics, we track 2000 electrons throughout the simulation. These electrons are chosen to be located inside the EV at $t = 150T_0$, with half of them in the forward-going current distributed throughout the structure (hereafter laser-induced current) and the other half in the return current. Their positions at $t = 150T_0$ and $t = 400T_0$ are represented by blue and red circles in Fig.~3(a), respectively. Most of the electrons in the laser-induced current stay within the EV, but the electrons in the return current are lost when they reach the end of the EV. This can also be seen from the representative orbits shown by the black curves.
%In other words, the majority of these two groups of electrons do not mix with each other,
% (because otherwise the forward-propagating electrons would also be lost from the EV if they enter the return currents)

\begin{figure}
\centering
\includegraphics[width=1\textwidth]{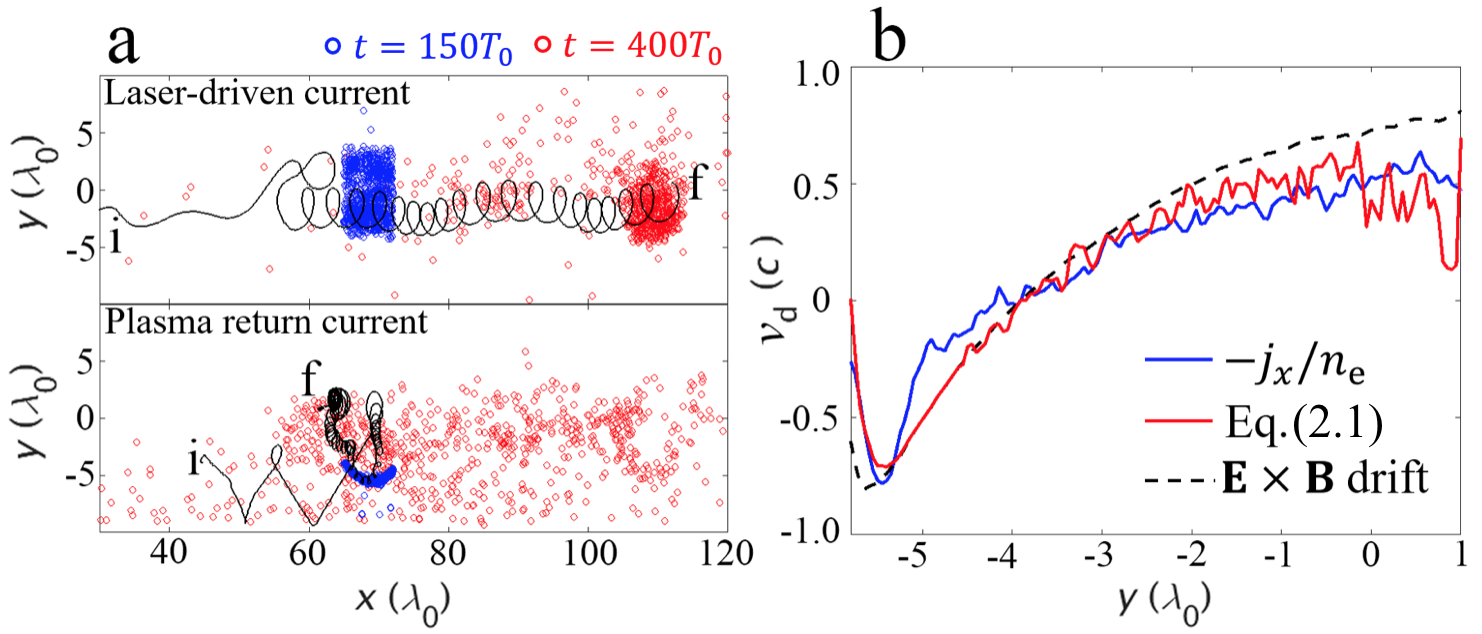}
\caption{The dynamics of the trapped electrons in the EV. (a) The upper and lower panel show the positions of 1000 tracked electrons (each) in the laser-driven current and return current, selected at $t = 150 T_0$. The red and blue circles show their position at $t = 150 T_0$ and $t = 400 T_0$, respectively. A representative electron trajectory in each case is plotted with black curve, i and f marks the initial and final points of the trajectory. (b) A comparison of the average drift velocity of electrons (blue line) and the prediction from Eq.~(\ref{eq1}) (red line), the black dashed line shows the $\mathbf{E}\times\mathbf{B}$ drift velocity [first term in Eq.~(\ref{eq1})]. All quantities are taken from the simulation at $t = 200 T_0$, along  $x = 83.5\lambda_0$.}
\label{fig.3}
\end{figure}

An individual electron drifts in the electric and magnetic fields self-generated by the vortex, where the drift velocity can be obtained by the sum of the $\mathbf{E}\times\mathbf{B}$ drift and grad-$B$ drift. (Note that only the electrons are magnetised, so the protons and ions are not subject to $\mathbf{E}\times\mathbf{B}$ drift in the time scale of interest)
\begin{equation} 
\frac{\mathbf{v_{\rm{d}}}}{c} = \frac{\mathbf{E}\times\mathbf{B}}{B^2} - \left(\gamma_{\rm{e}}-\frac{1}{\gamma_{\rm{e}}}\right)\frac{m_{\rm{e}}c^2\mathbf{B}\times\nabla B}{2eB^3},
\label{eq1}
\end{equation}
{\noindent}where $\gamma_{\rm{e}}$ is the relativistic factor of electrons. By assuming $\gamma_{\rm{e}} = \gamma_a \equiv \sqrt{1+a_0^2/2}$ in the laser-induced current and $\gamma_{\rm{e}}\sim 1.4$ in the relatively cold return current, the prediction of Eq.~(\ref{eq1}) matches the PIC simulation data as shown by Fig.~3(b). Moreover, it can be seen that the $\mathbf{E}\times\mathbf{B}$ drift is the dominating term in Eq.~(\ref{eq1}); the grad-$B$ drift only becomes important in the upper boundary of the EV, where the gyro-radii of electrons are not negligible compared to the variation scale of magnetic fields.\\

\section{Model of EV velocity in strongly non-uniform plasmas.}

Since the laser-induced electrons propagate with the EV [see Fig.~3(a) and Supplemental Movie], their average drift velocity must equal the EV propagation speed. 
Due to the approximate anti-symmetry of $E_x$, shown in Fig.~2,  the drifts in $y$ direction must average to zero and the EV moves in a direction consistent with $\mathbf{B} \times \nabla n$, as found in weakly inhomogeneous plasmas \citep{Richardson2013, Angus2014}. 

We model the EV  as two oppositely-travelling electron streams: the electrons going backwards are in a thin layer with high density, and the ones going forward are in a larger region but with significantly lower density, as shown in  Fig.~4(a). 
For simplicity, the electron and ion densities are taken to be uniform in the return current layer: $n_{\rm{e}} = n_{\rm{e}1}$ and $n_{\rm{i}} = n_{\rm{i}1}$, and  the $y$ coordinate is set to be zero at the bottom of the EV. The interface between the two oppositely-travelling currents is at $y = y_1$ and the upper boundary of the EV is at $y = y_2$ (where the gyro-radii of the electrons become comparable to the size of EV).
As the proton density is negligible, the positive charge density is $Zen_{\rm{i1}}\exp{[-(y-y_1)/\sigma]}$, where $Z$ is the charge number and $n_{\rm{i}1}$ is the immobile ion density at $y = y_1$. 
%Since the thickness of return current layer is much smaller than $\sigma$, it is reasonable to assume the electron and ion densities to be uniform in the layer: 
Furthermore, we assume the shape of EV to be elongated in the $x$ direction (which is increasingly well satisfied for high $a_0$), effectively reducing the problem to one dimension ($\partial/\partial x \ll \partial/\partial y$). 
The relevant electric and magnetic fields at $y_1$ are $E_y=E_1$, $B_z=B_1$, and $E_2$, $B_2$ at $y_2$, respectively, where $|B_2| \ll |B_1|$ and $|E_2| \ll |E_1|$.

\begin{figure}
\centering
\includegraphics[width=0.8\textwidth]{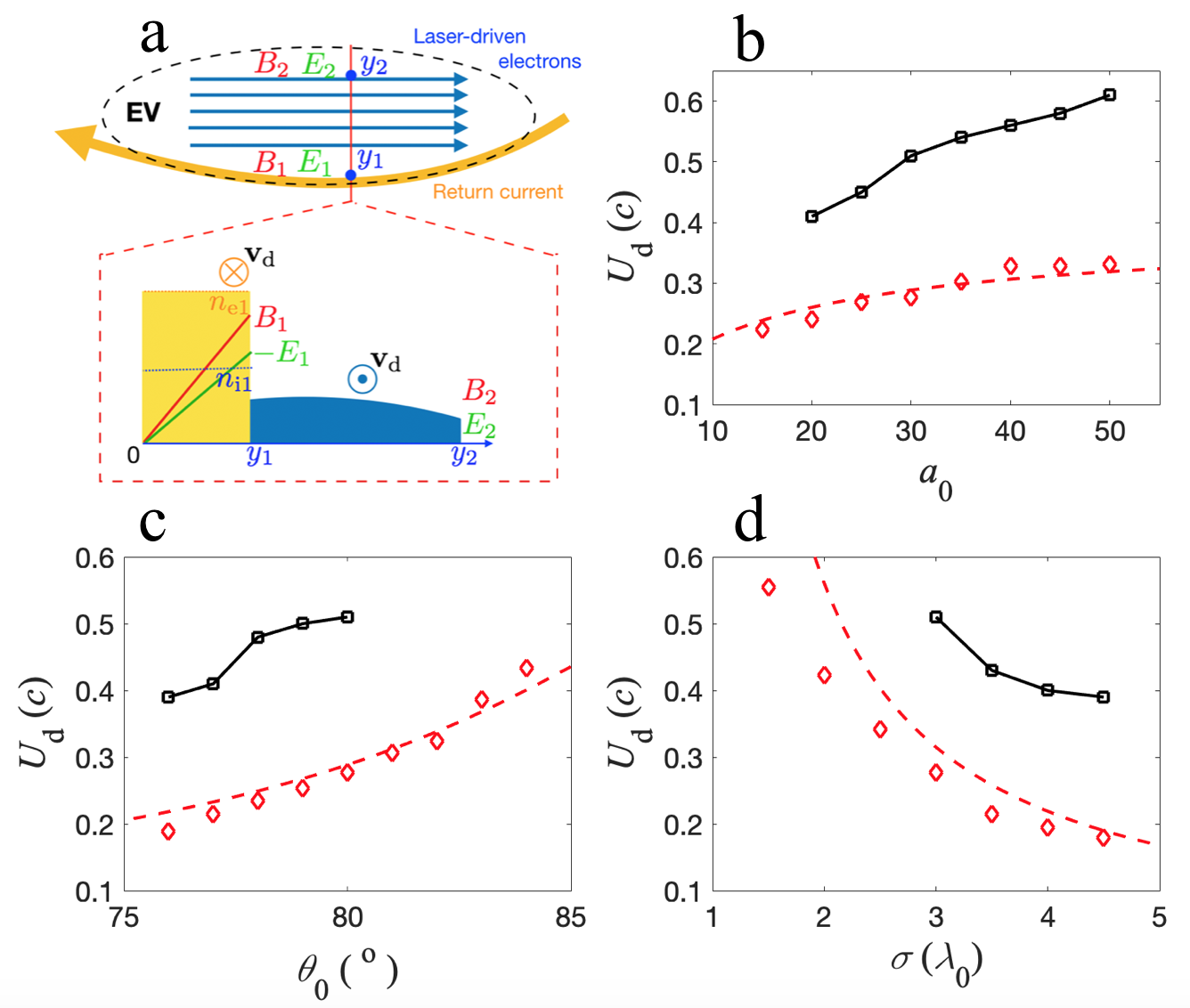}
\caption{Analytical model to calculate the EV drift velocity. (a) The 1D EV model with two oppositely-travelling electron streams. The drift velocity of EV and the maximum velocity of the accelerated protons (black squares) are plotted as  functions of (b) $a_0$, (c) $\theta_0$, and (d) $\sigma$, respectively. These parameters are set to their default values $a_0 = 30$, $\theta_0 = 80^{\circ}$, and $\sigma = 3\lambda_0$, when not being scanned.  The red diamonds and dashed lines in (b-d) represent the drift velocities in PIC simulations and the fit suggested by Eq.~(\ref{ud}).}
\label{fig.4}
\end{figure}

The electric and magnetic fields satisfy Maxwell's equations $\partial E_y/\partial y = 4\pi e(Zn_{\rm{i}}-n_{\rm{e}})$ and $\partial B_z/\partial y = -4\pi j_x/c \approx -4\pi en_{\rm{e}}v_{{\rm d} x}/c$. Integrating these two equations in the region of laser-induced current gives an estimate for the drift velocity of the EV
\begin{equation} 
U_{\rm{d}} = \frac{\int_{y_1}^{y_2}v_{\rm{d}}n_{\rm{e}}{\rm{d}}y}{\int_{y_1}^{y_2}n_{\rm{e}}{\rm{d}}y} \approx \left(\frac{E_1}{B_1}+\frac{4\pi eN_{\rm{i}}}{B_1}\right)^{-1}c,
\label{eq2}
\end{equation}
where $N_{\rm{i}} = Zn_{\rm{i}1}\sigma(1-\exp{[(y_1 - y_2)/\sigma])}$  
is the ion areal density inside the EV. 
Thus, the drift velocity of an EV depends
on $N_i$ as well as the electric and magnetic fields 
at $y_1$.

The first term in the bracket on the right hand side of Eq.~(\ref{eq2}) can be obtained by integrating Maxwell's equations in the return current layer (note that the $\mathbf{E}\times\mathbf{B}$ drift is the dominating term in this region, as illustrated in Fig.~3(b), so that $v_{\rm{d}}$ can be estimated by $cE_y/B_z$), which gives $E_1/B_1 = -\sqrt{1-Zn_{\rm{i}1}/n_{\rm{e}1}}$. From this it follows that  $|E_1/B_1|$ varies over a range of $\sim$ 0.1, that is typically an order of magnitude smaller than the variation of the second term in Eq.~(\ref{eq2}). As a representative value, we assume $Zn_{\rm{i}1}\approx n_{\rm{e}1}/2$, yielding $E_1/B_1\approx-0.7$.

An estimate of the second term in  Eq.~(\ref{eq2}) relies on the fact that the amplitude of $B_1$ in the EV approximately equals the magnetic field in the laser created channel (see Fig.~1). Consider the case where all the electrons in the channel move forward with velocity $\sim c$, and the bottom of the channel is at the position $y = y_1$. Then $B_1\approx 4\pi eZn_{\rm{i}1}\sigma[1-\exp(-R_{\rm{ch}}/\sigma)]$, where $R_{\rm{ch}} = \lambda_0\sqrt{\gamma_an_{\rm{c}}/4\pi^2n_{\rm{m}}}$ is the characteristic radius of the laser driven channel \citep{Borisov1995, Pukhov1999}, and $n_{\rm{m}}$ is the local plasma density corresponding to the maximum laser penetration depth in the NCD plasma.

From Snell's law, we obtain $\eta\sin\alpha = \eta_0\sin\theta_0$, where $\eta \approx 1 - n_{\rm{e}}/(2\gamma_a)$ is the refractive index of the plasma [the initial value is $\eta_0 \approx 1-n_0/(2\gamma_a)$, and $\alpha$ is the angle between the laser propagation direction and $y$ axis. Thus, by setting $\alpha = \pi/2$, the density corresponding to the maximum penetration of the laser is obtained as
%\begin{equation}
$n_{\rm{m}}/n_{\rm{c}} = 2\gamma_a - (2\gamma_a - n_0/n_{\rm{c}})\sin\theta_0$.
%\end{equation}

For large EVs ($N_{\rm{i}} \approx Zn_{\rm{i}1}\sigma$), the drift velocity in terms of laser-plasma parameters can then be approximated as
\begin{equation} 
U_{\rm{d}} \approx c \left[\frac{2\pi\sigma}{\lambda_0}\sqrt{2-\left(2-\frac{n_0/n_{\rm{c}}}{\sqrt{1+a_0^2/2}}\right)\sin\theta_0}-0.7\right]^{-1}.
\label{ud}
\end{equation}

The most important prediction of Eq.~(\ref{ud}) is that the drift velocity, and so the maximum proton energy, does not depend crucially on the laser intensity, since the plasma density at the incident point is typically $n_0/n_{\rm{c}}\ll a_0$. Therefore, controlling the laser incident angle and the plasma density gradient can be effectively used to adjust the proton energy. This agrees well with the PIC simulations as shown by Fig.~4(b-d), $U_{\rm{d}}$ depends strongly on the laser incident angle and plasma scale length, and weakly on the laser intensity (the scaling flattens further as $n_0$ decreases). 

The maximum proton speed obtained in the PIC simulations are overlaid on the plot in
Fig.~4(b-d) with black squares, and they are about twice the corresponding drift velocity. 
There are cases where no accelerated proton energy is plotted above $U_{\rm{d}}$. This indicates that protons cannot be captured by the EV for those parameters, either because the acceleration field $E_x$ is too weak [in Fig.~4(b) when $a_0 = 10$] or because the EV is drifting too fast (for the cases when $U_{\rm{d}}>0.4c$). In this case one could increase the values of $a_0$, $\theta_0$, or $\sigma$ as suggested in Fig.~4(b-d).

Unlike most other proton acceleration mechanisms, which require high laser intensity to reach high ion
energies, the scheme proposed here provides accessible degrees of
freedom to control the acceleration process, allowing accelerating
protons to hundreds of MeV with a narrow energy spread. 
According to Fig.~4 the laser parameters required to produce a 
100-200 MeV monoenergetic, self-injected proton beam are 
%$10^{21}$ W/cm$^2$ intensity, 300 TW power, and 6 J energy. These parameters are 
within reach of existing laser facilities.  
% the laser $a_0$ to ensure self-injection (without reducing the maximum acceleration energy) 

\section{Discussion}

The grazing incidence of the laser pulse allows the laser to accelerate hot electrons parallel to the target surface, the energy in the resulting coherent motion can subsequently be transferred to protons during the EV-acceleration stage. If the incident angle is too small, the EV will be depleted at a premature state of the proton acceleration. 

\begin{figure}
\centering
\includegraphics[width=1\textwidth]{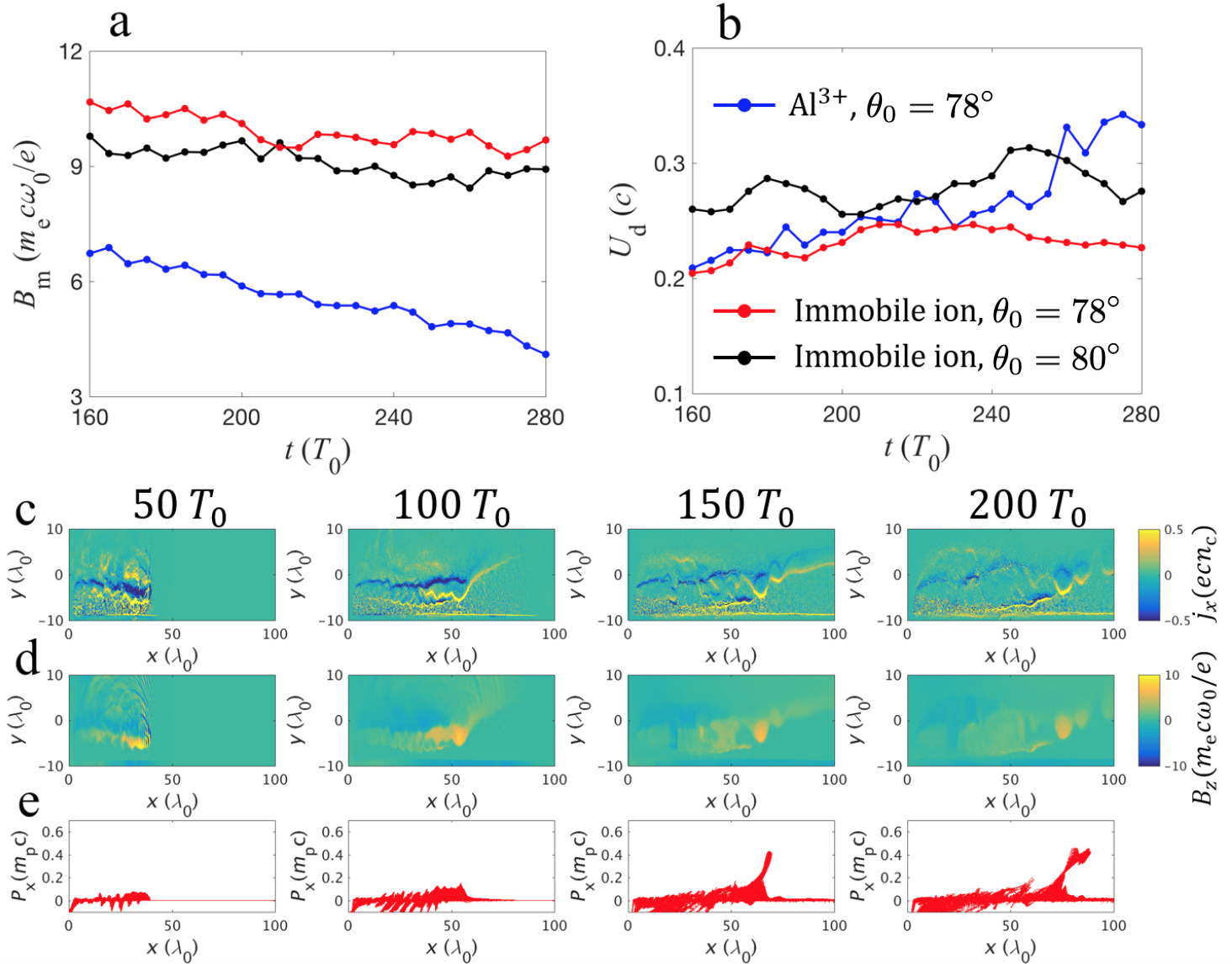}
\caption{The effect of dynamical background heavy ions. Time evolution of (a) the maximum magnetic field and (b) the EV drift velocity. The blue line corresponds to a simulation with mobile Al$^3+$ background ions and a laser incidence angle of 78$^{\circ}$. The red and black lines show  cases with immobile ions, and angles 78$^{\circ}$ and 80$^{\circ}$, respectively.
Snapshots of (c) the current density $j_x$, (d) the out-of-plane magnetic field $B_z$, and (e) the $x$--$P_x$ proton phase space at $t = 50T_{0}, 100T_{0}, 150T_{0}$ and $200T_{0}$, respectively, in a simulation with a mobile background heavy ion species, Al$^{3+}$.}
\label{fig.5}
\end{figure}

Also, when the incident angle of the laser is less than some critical angle (so that the plasma density on both sides of the laser becomes significantly higher than in the electron depletion region created by the laser in the middle), the laser self-channeling effect becomes important, and the laser-driven electrons will be trapped in the channel running towards the high-density region. In this situation no EV forms. Therefore, we only consider the grazing incident angle, above 75$^{\circ}$, in this work.\\

\begin{figure}
\centering
\includegraphics[width=1\textwidth]{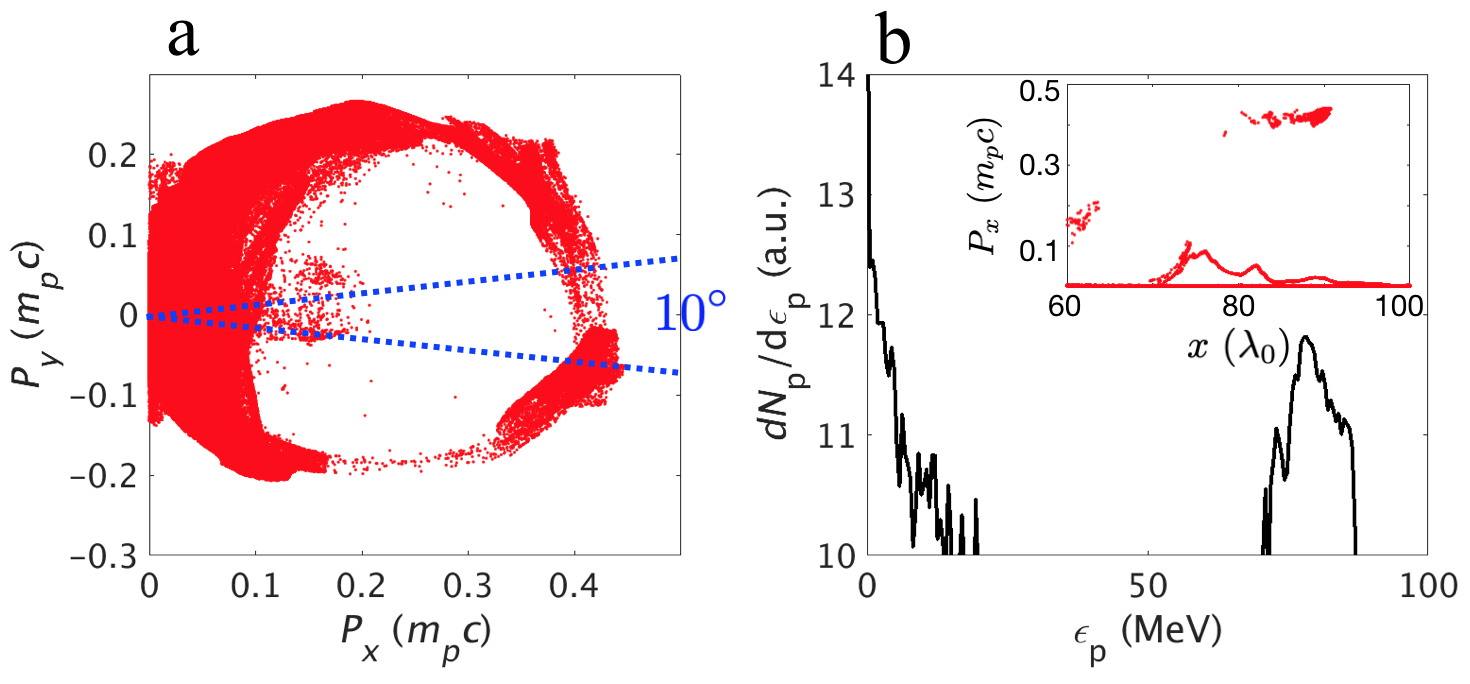}
\caption{Accelerated protons in a simulation with mobile ions. (a) Proton distribution in the $P_x$--$P_y$ phase space, and (b) proton energy spectrum within an opening angle of $10^{\circ}$ at $t = 200 T_0$. The inset in (b) show the $x$--$P_x$ phase space map of these protons.}
\label{fig.6}
\end{figure}

It should be noted that the immobile ion background assumption used in the simulation of Fig.~1 is reasonable, as the required low charge-to-mass ratio (i.e.~low degree of ionization) of the heavy ions can be well satisfied in the region where the EV propagates and the ions are being accelerated. The high intensity laser light that can ionize the plasma to a higher degree never reaches this region, as it is fully depleted by the time the vortex is fully formed. To illustrate the effect of dynamical background heavy ions, here we present a simulation where we replace the immobile ions with Al$^{3+}$. The initial ratio of Al$^{3+}$ and H$^{+}$ in the expanding near-critical-density plasma is assumed to be 18:1. In addition, the laser incident angle is $\theta_0 = 78^{\circ}$ (as motivated below), all other parameters are unchanged from those in Fig.~1.

As seen in Fig.~5, the EV is slowly dissipated as ions are scattered out of the vortex by its electrostatic field. As a result, the EV loses energy and its lifetime is reduced. Figure 5(a) shows that the maximum magnetic field in the EV is reduced by 40$\%$ within 120 $T_0$ in the mobile background ion case (Al$^{3+}$), while in the immobile ion case, the reduction is only about 10$\%$. The decrease in magnetic field results in a slight increase in the EV drift velocity as shown in Fig.~5(b). This can be understood from Eq.~(\ref{eq2}), $U_{\rm{d}}\sim B_1/N_{\rm{i}}\sim \sigma\sqrt{n_{\rm{m}}/\gamma_a}$, and $B_1\sim\sqrt{\gamma_a n_{\rm{m}}}$, which yields $U_{\rm{d}}\sim\gamma_a/\sigma B_1$. Because of this, there is no analogous spectral broadening effect as in shock wave acceleration \citep{Macchi2012}. 

As shown in Fig.~5(e), the qualitative features of the process and the overall dynamics remain the same with mobile ions. Importantly, the fields of the EV remain sufficiently strong during the timescale needed for proton acceleration ($\sim 50T_0$). %Thus the scheme proposed in the paper is useful as a particle accelerator.
To compensate for the somewhat weaker fields in the EV, the incident angle of the laser pulse has been reduced. To ensure proton self-injection, the maximum incident angle is $\theta_0 = 78^{\circ}$, which leads to an acceleration energy of $\sim 80$ MeV. This is approximately 60$\%$ of the value obtained in the immobile ion case. Otherwise, the proton phase space and energy spectrum are similar to those with immobile ions, as shown in Fig.~6. These results suggest that heavier background ion species (that tend to have smaller charge-to-mass ratio due to partial ionization) are preferable for achieving a more energetic proton beam with longer duration.\\

\begin{figure}
\centering
\includegraphics[width=1\textwidth]{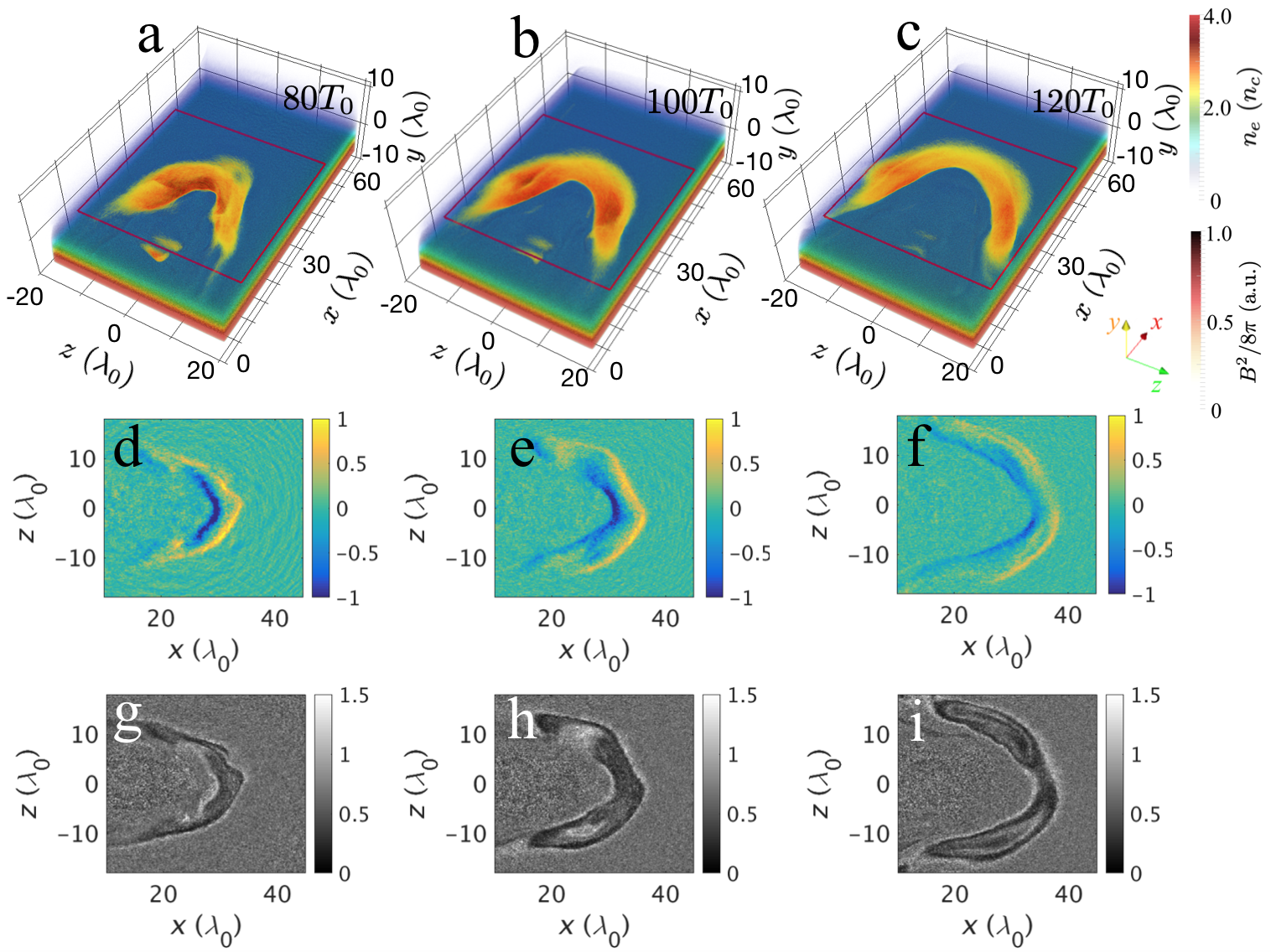}
\caption{The 3D structure of the electron vortex. Magnetic energy density (white-orange colour scale) and the electron density (rainbow-colour scale) in the 3D PIC simulation at (a) $t = 80T_{0}$, (b) $100T_{0}$, and (c) $120T_{0}$. (d-f) and (g-i) show the $x$ component of the electric field, and the electron density depletion for the cross-section at $y = -1.6\lambda_0$ [for regions marked by the red rectangle in (a-c)].}
\label{fig.7}
\end{figure}

Finally, we present a 3D simulation of the considered setup. We use a circularly-polarised laser with a reduced peak intensity $I_0 = 3\times10^{20}$ W/cm$^{2}$ and an elliptical focal spot ($I = I_0\exp(-y^2/w_y^2 - z^2/w_z^2)$, where $w_y = 3\lambda_0$ and $w_z = 9\lambda_0$). The initial plasma density profile is assumed to be uniform in the $z$ direction, the laser incident angle $\theta_0 = 75^{\circ}$, and other parameters are the same as quoted for the 2D simulation of Fig~1. The resolution needed to be reduced compared to 2D, for computational feasibility: the simulation box is $x\times y \times z= 100\lambda_{0} \times 20\lambda_{0} \times 40 \lambda_{0}$, which is sampled by $1000 \times 200 \times 400$ cells with 3 macro particles per cell for both electrons and ions. Circular polarization is chosen as it can generate a higher current density, leading to a stronger EV. The linearly polarised laser is known to produce fast electron ``jets", twice per laser cycle, which appears as an unwanted loss mechanism. An elliptical laser focus spot is employed to elongate the flux rope, thereby mitigating the leakage of electrons along the field lines, and a lower laser incident angle is used to ensure proton injection. 

Figure 7 illustrates that the electron vortex has a ``magnetic flux rope" structure in 3D space, where the magnetic field is highly concentrated in a flux tube that lays in the $x$-$z$ plane and resembles a horseshoe. The plasma density depletion and charge separation fields are similar to those in the 2D simulations, and the structure stays stable for more than 40 $T_0$. The formation and drifting process of the EV in 2D and 3D simulations are essentially the same, which indicates the robustness of the underlying physics, i.e., the electrons trapped in the vortex are subject to an $\mathbf{E}\times\mathbf{B}$ drift which drives the collective drift motion of the vortex structure; the strong magnetic pressure leads to charge separation in the vortex and a population of protons along its path are captured and accelerated (see Fig.~8).

\begin{figure}
\centering
\includegraphics[width=1\textwidth]{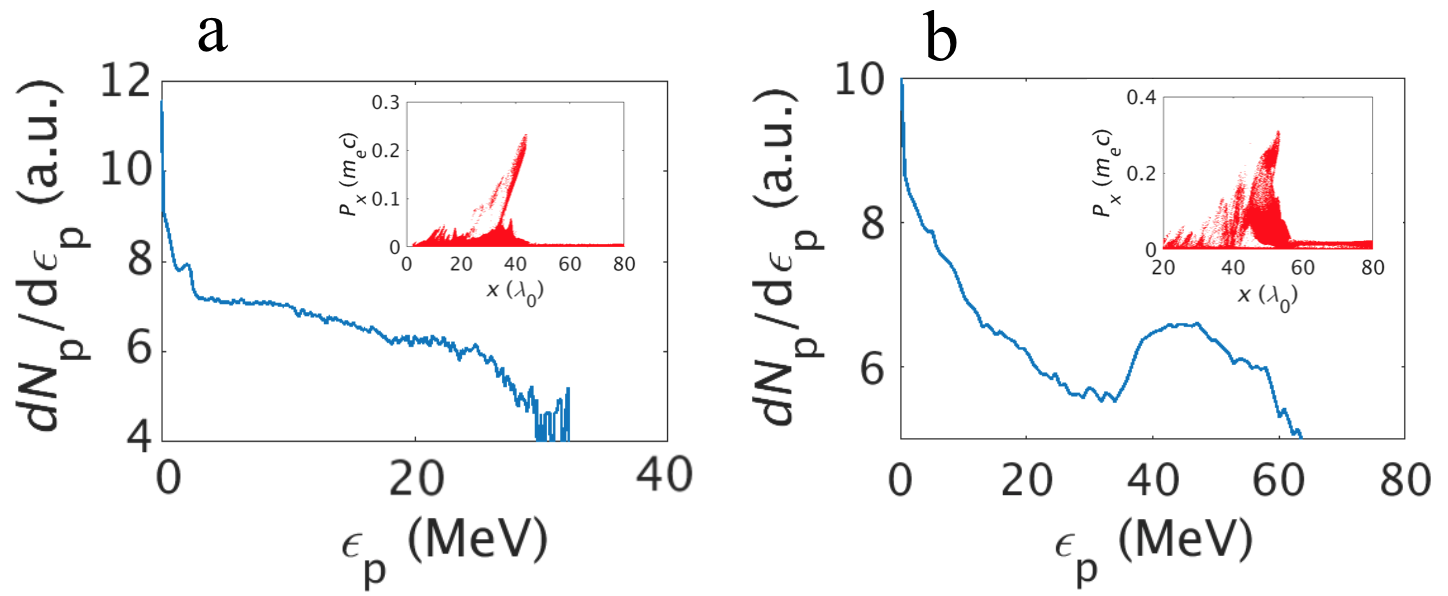}
\caption{Accelerated proton spectra in 3D simulations within opening angle $|\theta_y| = {\rm{arctan}}(|P_y|/P_x) < 10^{\circ}$ at simulation time $t = 140 T_0$. (a) Laser peak intensity $I_0 = 3\times10^{20}$ W/cm$^{2}$, with elliptical focus spot$w_y = 3\lambda_0$ and $w_z = 9\lambda_0$, incident at $\theta_0 = 75^{\circ}$. (b) ``Quasi-2D" case, with $w_z \gg w_y$, laser peak intensity $I_0 = 5.5\times10^{20}$ W/cm$^{2}$, incident at $\theta_0 = 78^{\circ}$}
\label{fig.8}
\end{figure}

However, in 3D the electrons can escape along the magnetic field lines in the third dimension (``leaking" effect), leading to a reduction of the current and the magnetic energy density. As a result, the lifetime of the electron vortex is expected to be shorter, usually resulting in a lower energy of the accelerated protons. As shown in Fig.~8(a), a monotonically decreasing proton spectrum is produced, with a cut-off energy of around 30 MeV, and the proton number (above 10 MeV) per solid angle is estimated to be $10^4$/MeV/msr. The lack of the high energy peak is caused by the rapid decay of the EV due to electron leakage along field lines. That the EV loses energy, and its drift velocity changes as it rises towards the lower-density region, happens within the time scale of the proton acceleration, resulting in a different spectral shape and low proton energies. The unwanted effects of the leakage can be mitigated by increasing the intensity and the ellipticity of the laser spot-shape. As an illustration, in Fig.~8(b) we show a case with an extremely elongated laser spot, $w_z\gg w_y$ [i.e.~$I = I_0\exp(-y^2/w_y^2)$], where a laser of intensity $I_0 = 5.5\times10^{20}$ W/cm$^{2}$ is incident at $\theta_0 = 78^{\circ}$, recovering the high energy proton peak in the spectrum. A quantitative study on the differences between 2D and 3D cases, induced by hot electron leakage, is computationally very expensive; the detailed optimization of the presented scheme in 3D is left for future studies.

%A quantitative study on the differences between 2D and 3D cases, induced by this effect, is computationally very expensive, as the extra degrees of freedom, such as the ellipticity of the laser spot-shape, should be taken into account to find the optimal parameter range for proton acceleration. The detailed optimization of the presented scheme in 3D is left for future studies.

\section{Conclusions}

In conclusion, we show that when a laser pulse irradiates an NCD plasma in a grazing angle, in the presence of a sharp, pre-formed target-normal density gradient, the resulting intense electric currents form an electron vortex after the laser depletion. The magnetic pressure leads to a significant reduction in the electron density in the centre of the vortex, and the electric field created by the charge separation can accelerate ions. The drift velocity of the EV can be effectively tuned by the incident angle of the laser and the density gradient of the plasma. Using PIC simulations we demonstrate that protons initially at rest can be captured and accelerated to maximum twice of the EV drift velocity with a narrow energy spread. 

\section*{Acknowledgments}
The authors acknowledge fruitful discussions with S Newton, J Martins, E Siminos, J Ferri, I Thiele and the rest of the PLIONA team. This work is supported by the Knut and Alice Wallenberg Foundation, the European Research Council (ERC-2014-CoG grant 647121), the National Natural Science Foundation of China (Grant No.11505262), the Ministry of Science and Technology of the People's Republic of China (Grant Nos. 2016YFA0401102 and 2018YFA0404803), and the Strategic Priority Research Program of the Chinese Academy of Sciences (Grant No. XDB16). Simulations were performed on resources at Chalmers Centre for Computational Science and Engineering (C3SE) provided by the Swedish National Infrastructure for Computing (SNIC).

\bibliographystyle{jpp}
\bibliography{references} 

\end{document}